\documentclass[12pt]{article}
\usepackage[cp1251]{inputenc}  
\usepackage[english]{babel}
\usepackage{amsmath,amssymb,epsfig,graphicx,doc,latexsym,amsfonts,bm,longtable}
\usepackage[bookmarksopen,colorlinks]{hyperref}
\usepackage[monochrome]{color}
 \hypersetup{colorlinks=true}

\begin{document}
\begin{center}
{\textbf{\LARGE{Three modes of the shear-flow-driven ion cyclotron
instability}}}

\bigskip
\bigskip
{\large{D. V. Chibisov}},{\large{V.S. Mikhailenko} and K.N.
Stepanov}
\end{center}

\bigskip

Kharkov National University, 61108 Kharkov, Ukraine

\begin{abstract}
The electrostatic shear-flow-driven ion cyclotron instability of
magnetic field aligned sheared plasma flow is investigated
analytically. It is shown that the shear-flow-driven electrostatic
ion cyclotron instability can be considered as a combination of
three different instabilities determined by theirs own mechanism of
excitation: ion-kinetic, ion-hydrodynamic and electron-kinetic. Each
of these instabilities are dominant in different ranges of the
wavelength along the magnetic field. The linearized dispersion
equation is solved within the limits of long and short waves along
the magnetic field where effects of electrons and ions,
respectively, are dominant in the development of these
instabilities.  The general criterion of instability excitation,
which couples the flow velocity shear and wave number across the
magnetic field, is obtained.

\end{abstract}

\section{INTRODUCTION}
The ion cyclotron instabilities of magnetic field-aligned sheared
plasma flows have been intensively investigated for the last two
decades [1-11]. Interest to
this problem was caused by the attempts to explain the anomalously
strong ion heating in a number of the satellites' observations where
the correlation in place and time of the broadband, low--frequency
waves, and transverse ion energization was detected at the
boundaries of plasma structures [12-16]. As it turned out, the shear flow
of ions along the magnetic field, together with the magnetic
field-aligned electron current and others mechanisms is a possible
source of free energy for the development of EIC instabilities. It
was suggested in Ref. \cite{Belova} that the ion cyclotron waves in
the magnetopause are excited by the magnetic-field aligned ion flows
with a transverse velocity gradient. In Ref.\cite{Belova}, a
dispersion equation for electrostatic ion cyclotron (EIC) waves in
plasma shear flows has been obtained and solved numerically in the
long wavelength limit along the magnetic field. It was shown that
these oscillations have short waveslength in the transverse to the
magnetic field direction. Ganguly \textit{et
al.} [2-4] have prowed that ion cyclotron
instability may be developed due to the combined effect of the
inverse ion cyclotron damping and velocity shear of the magnetic
field-aligned ion flow. It was obtained in
Ref. [2-4] that different cyclotron
harmonics have approximately the same excitation conditions, which
explains the presence in the ion cyclotron waves spectrum in the
Earth's auroral magnetosphere numbers of cyclotron harmonics. By
using a hydrodynamic treatment, Merlino \cite{Merlino} has shown the
possibility of excitation of ion cyclotron instability due to shear
of the ion flow. Mikhailenko \textit{et
al.}\cite{Mikhailenko06,Mikhailenko07} using the kinetic approach,
have found that the shear flow of ions could lead to the splitting
of EIC oscillations onto two modes. One of them is the
shear-modified EIC mode, which also exists in plasma flows without
velocity shear and can be unstable due to the field-aligned electron
current. The second mode is caused by the shear flow of ions and can
be unstable even in a current-free plasma due to combined effect of
the velocity shear and inverse of electron Landau damping. It was
obtained in Refs.\cite{Mikhailenko06,Mikhailenko07}, that EIC waves
may be excited by the hydrodynamic mechanism, if the value of
velocity shear flow and wave numbers satisfy the certain conditions.
It was shown in Refs.\cite{Merlino, Mikhailenko06, Mikhailenko07},
that velocity shear-induced ion cyclotron instability in the limit
of long wavelength along the magnetic field becomes similar to an
aperiodic instability considered by D'Angelo\cite{Angelo} and is, in
fact, the extension of this instability onto the ion cyclotron
frequency range. The excitation of shear-flow-driven EIC instability
in a collisional plasma, which is the signature of the bottomside
ionosphere was studied by Mikhailenko \textit{et
al.}\cite{Mikhailenko08}. That instability is excited due to
electron-neutral collisions in the limit of long wave length of the
unstable waves along the magnetic field, for which effect of the
electron inverse Landau damping is negligible. Kaneko \textit{et
al.} \cite{Kaneko} have carried out the three-dimensional
electrostatic particle simulations and shown the possibility of
shear-flow-driven EIC instability. Simultaneously, a number of
studies (see for example Refs.\cite{Teodorescu,Kim}) attempted to
detect in the Q - machine the shear-flow-driven EIC instability.
However, as stated in Ref.\cite{Kim} this instability was not
observed yet.

In this paper, we analyze  the excitation mechanisms of the
shear-flow-driven EIC instabilities in a collisionless currentfree
sheared plasma.  We find that shear-flow-driven ion cyclotron
instability is developed by different mechanisms of excitations,
which dominate in certain ranges of the wavelength along the
magnetic field. According to that we find reasonable to define three
different modes of that instability: ion-kinetic mode,
ion-hydrodynamic mode and electron-kinetic mode. The regions of the
of the wavelength along the magnetic field, where these modes are
dominate and corresponding growth rates are found here. We have
obtained also the criterion of shear-flow-driven EIC instabilities
for the different ion cyclotron harmonics, which connects the flow
velocity shear and wave numbers across the magnetic field.

The paper is organized as follows. Sec. 2 is devoted to the analysis
of the shear-flow-driven EIC instabilities of the fundamental ion
cyclotron harmonic. In Sec. 3 the properties of these instabilities
for an arbitrary high number of the ion cyclotron harmonic are
considered. Conclusions are given in Sec. 4.
\section{THE INSTABILITY OF THE FIRST CYCLOTRON HARMONIC}

The start with dispersion equation for the oscillations in the ion
cyclotron frequency range in homogeneous magnetic-field aligned
plasma shear flow, which is given by\cite{Ganguli,Mikhailenko06}
\begin{eqnarray}
\varepsilon\left({\mathbf{k},\omega
}\right)=&&1+\frac{1}{k^2\lambda_{De}^2}\left(1+i\sqrt{\pi}z_{e0}
\left(z_{e0}\right)\right)+\frac{1}{k^2\lambda_{Di}^2}
\left[1-\frac{k_{y}}{k_{z}}S_{i} + i\sqrt{\pi}\sum\limits_{n
=-\infty}^{\infty}W\left(z_{i n}\right)
A_{n}\left(b_{i}\right)\right.\nonumber\\
&&\left.\times\left(\frac{\omega-k_{z}V_{0i}}{\sqrt{2}k_{z}v_{Ti}}
-\frac{k_{y}}{k_{z}}S_{i}z_{in}\right)\right]=0,\label{1}
\end{eqnarray}
where  $\lambda_{D\alpha}$ is the Debye length,
$A_{n}\left(b_{i}\right)=I_{n}\left(b_{i}\right)e^{-b_{i}}$,
 $I_{n}$ is the modified Bessel function, $b_{i}= k_\bot^2 \rho_{Ti}^2$,
$\rho_{Ti}= v_{Ti}/\omega_{ci}$  is the thermal Larmor radius,
$S_{i}=dV_{0i}\left(X\right)/\omega_{ci}dX$ is the normalized flow
velocity shear, $z_{\alpha n} = \left(\omega-n\omega_{c\alpha} -
k_{z}V_{0\alpha}\right)/\sqrt{2}k_{z}v_{T\alpha}$ and
$W\left(z\right)=e^{-z^{2}}\left(1+\left(2i/\sqrt{\pi}
\right)\int\limits_{0}^{z}e^{t^{2}}dt\right)$.

We consider first the instability of the fundamental cyclotron
harmonic, assuming that
$\omega\left(\mathbf{k}\right)=\omega_{ci}+k_{z}V_{0i}+\delta\omega
\left(\mathbf{k}\right)$ with
$\delta\omega\left(\mathbf{k}\right)\ll \omega_{ci}$. Assume that
the $z_{in}$ argument of the $W$ - function in the sum over
cyclotron harmonics has an arbitrary value for the fundamental
harmonic, while $|z_{in}|>1$ in the remaining sum. That is valid
when the inequality $k_{z}\rho_{Ti}<1$ is satisfied. Using the
asymptotic form for $W$ - function for large argument values,
$W(z_{i})\thicksim
e^{-z^{2}_{i}}+\left(i/\sqrt{\pi}z_{i}\right)\left(1+1/2z_{i}^{2}\right)$,
we carry out the summation over the cyclotron harmonics for $n\neq1$
\begin{eqnarray}
\sum\limits_{n \neq1}W\left(z_{i
n}\right)A_{n}\left(b_{i}\right)\left(\frac{\omega-k_{z}V_{0i}}
{\sqrt{2}k_{z}v_{Ti}}-\frac{k_{y}}{k_{z}}S_{i}z_{in}\right)\approx-
\frac{1-A_{0}\left(b_{i}\right)}{b_{i}}-\frac{k_{y}}{k_{z}}S_{i}
\left(1-A_{0}\left(b_{i}\right)\right).\label{2}
\end{eqnarray}
Then dispersion relation (\ref{1}) reduces to the form
\begin{eqnarray}
k^2\lambda_{Di}^2\varepsilon\left({\mathbf{k},\omega
}\right)&&=1+\tau\left(1+i\sqrt{\pi}z_{e0}W\left(z_{e0}\right)\right)
-\frac{1-A_{0}\left(b_{i}\right)}{b_{i}}-\frac{k_{y}}
{k_{z}}S_{i}A_{1}\left(b_{i}\right)\nonumber\\
&& +i\sqrt{\pi}W\left(z_{i
1}\right)A_{1}\left(b_{i}\right)\left(\frac{\omega_{ci}}
{\sqrt{2}k_{z}v_{Ti}}+z_{i1}-\frac{k_{y}}{k_{z}}S_{i}z_{i1}\right)=0.\label{3}
\end{eqnarray}
We use in what follows the normalized wavelength along the magnetic
field, $\lambda=1/k_{z}\rho_{Ti}$, instead of variable $k_{z}$.
Considering $z_{i1}$ as the normalized complex frequency, we find
the solution $z_{i1}\left(\lambda\right)$ of the Eq. (\ref{3}) for
EIC instability in the short wavelength limit, at which the
electrons are adiabatic with $z_{e0}\ll1$ and instability is
developed due to the inverse of ion cyclotron damping, and in long
wavelength limit, at which ion cyclotron damping is negligible and
EIC instability is developed due to inverse electron Landau damping.

We find first the short wavelength threshold. It is important to
note, that in shearless plasma flow the ion cyclotron instability
does not developed in this limit. The threshold values for variables
$\lambda$ and $z_{i1}$ we obtain by equating to zero the real and
imaginary parts of Eq. (\ref{3}),
\begin{eqnarray}
\left\{\begin{array}{ll}\lambda/\sqrt{2}+z_{i1}-\lambda k_{y}\rho_{Ti}
S_{i}z_{i1}=0,
\\ 1+\tau-\left(1-A_{0}\left(b_{i}\right)\right)/b_{i}-\lambda k_{y}
\rho_{Ti}S_{i}A_{1}\left(b_{i}\right)=0.\end{array} \right.\label{4}
\end{eqnarray}
This set has a solution when inequality $k_{y}\rho_{Ti}S_{i}>0$ is
met. For this case we obtain the short-wavelength threshold value
$\lambda_{1s}$ for the excitation of the instability, as well as the
threshold value of the normalized complex frequency $z_{1s}$, which
is the real at that threshold,
\begin{eqnarray}
\lambda_{1s}&&=\frac{1}{k_{y}\rho_{Ti}S_{i}A_{1}\left(b_{i}\right)}
\left(1+\tau-\frac{1-A_{0}\left(b_{i}\right)}{b_{i}}\right),\label{5}
\\z_{1s}&&=\frac{1}{\sqrt{2}S_{i}k_{y}\rho_{Ti}}\left(1+\frac{A_{1}
\left(b_{i}\right)}{1-G_{1}+\tau}\right),\label{6}
\end{eqnarray}
where
$G_{1}=A_{1}\left(b_{i}\right)+\left(1-A_{0}\left(b_{i}\right)\right)/b_{i}$
and index $s$ means the short-wavelength instability threshold of
the first cyclotron harmonic. The value
$\delta\omega=\delta\omega_{01}$ at the instability threshold is
\begin{eqnarray}
\delta\omega_{01}=\frac{\omega_{ci}A_{1}\left(b_{i}\right)}{1-G_{1}+\tau},\label{7}
\end{eqnarray}
which coincides with value of $\delta\omega_{01}$ in shearless flow.
The approximate solution to Eq.(\ref{3}) for $z_{i1}$ at the
vicinity of instability threshold we obtain by Taylor series
expansion of Eq.(\ref{3}) in powers of
$\left(\lambda-\lambda_{1s}\right)$, with zero-order and linear
terms retained,
\begin{eqnarray}
z_{i1}\simeq
z_{01}+z'_{\lambda}\left(\lambda_{1s}\right)\left(\lambda-\lambda_{1s}\right).\label{8}
\end{eqnarray}
Here $z'_{\lambda}\left(\lambda_{1s}\right)=-\varepsilon'_{\lambda}/\varepsilon'_{z}$ with
\begin{eqnarray}
k^2\lambda_{Di}^2\varepsilon'_{\lambda}\left(z_{1s}\right)=-i\sqrt{\pi}W\left(z_{
1s}\right)A_{1}\left(b_{i}\right)\frac{z_{1s}}{\lambda_{1s}}-
k_{y}\rho_{Ti}S_{i}A_{1}\label{9}
\end{eqnarray}
and
\begin{eqnarray}
k^2\lambda_{Di}^2\varepsilon'_{z}\left(z_{1s}\right)=-i\sqrt{\pi}W\left(z_{
1s}\right)A_{1}\left(b_{i}\right)\frac{\lambda_{1s}}{\sqrt{2}z_{1s}}.\label{10}
\end{eqnarray}
The dispersive part,  $\delta\omega$, of the ion cyclotron wave
frequency and the growth rate at the vicinity of the instability
threshold can be obtained from Eq. (\ref{6}) as
\begin{eqnarray}
\delta\omega\simeq\delta\omega_{01}\frac{\lambda_{1s}}{\lambda}+&&\delta\omega_{01}
\left(\frac{\sqrt{2}k_{y}\rho_{Ti}S_{i}\text{Im}W\left(z_{
1s}\right)}{\sqrt{\pi}|W\left(z_{
1s}\right)|^{2}}-\frac{A_{1}\left(b_{i}\right)}{1-G_{1}+\tau}\right)
\left(1-\frac{\lambda_{1s}}{\lambda}\right),\label{11}
\\\gamma&&\simeq\delta\omega_{01}\frac{\sqrt{2}k_{y}\rho_{Ti}S_{i}\text{Re}W\left(z_{
1s}\right)}{\sqrt{\pi}|W\left(z_{
1s}\right)|^{2}}\left(1-\frac{\lambda_{1s}}{\lambda}\right).\label{12}
\end{eqnarray}
Because
$\gamma\propto\text{Re}W\left(z_{1s}\right)=exp\left(-z_{1s}^{2}\right)$,
the instability growth rate is exponentially small at the vicinity
of the threshold for the values of velocity shear flow and
transverse wave number such that $\sqrt{2}k_{y}\rho_{Ti}S_{i}<1$ and
it is not exponentially small with opposite inequality
$\sqrt{2}k_{y}\rho_{Ti}S_{i}>1$  i. e. for larger values of the
velocity shear and transverse wave numbers. The magnitude of the
growth rate is affected also by the factor
$k_{y}\rho_{Ti}A_{1}\left(b_{i}\right)$; the growth rate decreases
as $\left(k_{y}\rho_{Ti}\right)^{3}$ for long waves with
$k_{y}\rho_{Ti}<1$, whereas the growth rate varies slightly for the
waves with $k_{y}\rho_{Ti}\gtrsim1$.

Thus, the EIC instability at the vicinity of the short wave
threshold is resulted from the combined effect of the inverse of ion
cyclotron damping velocity shear and is the ion-kinetic mode of the
shear-flow-driven EIC instability. The most unstable waves are those
with transverse wave numbers $\sqrt{2}k_{y}\rho_{Ti}S_{i}>1$ and
$k_{y}\rho_{Ti}\gtrsim1$. In addition, since the development of this
mode is caused  by the thermal motion of ions along the magnetic
field, this instability will be resulted in the longitudinal heating
of ions.

The growth rate of the ion-kinetic mode of the shear-flow-driven EIC
instability increases with an increase of wavelength along the
magnetic field. Simultaneously, the value $\left|z_{i1}\right|$ also
increases. If the transverse wave numbers meet the inequality
$\sqrt{2}k_{y}\rho_{Ti}S_{i}<1$, the growth rate, remaining
exponentially small quantity, reaches a maximum at a certain value
$\lambda$ and then rapidly decreases. If the transverse wave numbers
meet the inequality $\sqrt{2}k_{y}\rho_{Ti}S_{i}>1$, the ion-kinetic
mode with an increase $\left|z_{i1}\right|>1$ turn into the
ion-hydrodynamic mode. The dispersion equation for the
ion-hydrodynamic mode of EIC shear-flow-driven instability can be
obtained from Eq. (\ref{3}) by using the asymptotic form of $W$ -
function for a large magnitudes of $|z_{i1}|>1$
\begin{eqnarray}
1+\tau\left(1+i\sqrt{\pi}z_{e0}W\left(z_{e0}\right)\right)-G_{1}
-\frac{A_{1}\left(b_{i}\right)}{\sqrt{2}z_{i1}}\lambda+
\frac{A_{1}\left(b_{i}\right)}{2z_{i1}^{2}}k_{y}\rho_{Ti}S_{i}\lambda=0,\label{13}
\end{eqnarray}
where $z_{e0}=\lambda/\sqrt{2\mu}$ and $\mu=m_{i}/m_{e}$. The
solution of Eq.(\ref{12}) is
\begin{eqnarray}
 && z_{i1}=\left[\lambda
A_{1}\left(b_{i}\right)r\pm\sqrt{\left(\lambda
A_{1}\left(b_{i}\right)\right)^{2}-4\left(1+\tau\left(1+i\sqrt{\pi}z_{e0}W\left(z_{e0}\right)\right)
-G_{1}\right)S_{i}\lambda
k_{y}\rho_{Ti}A_{1}\left(b_{i}\right)}\right]
\nonumber\\
&&\times\left[2\sqrt{2}k_{y}\rho_{Ti}\left
(1+\tau\left(1+i\sqrt{\pi}z_{e0}W\left(z_{e0}\right)\right)-G_{1}\right)\right]^{-1}.\label{14}
\end{eqnarray}
The long wavelength threshold for ion-hydrodynamic mode we find from
Eq.(\ref{10}) assuming
$\text{Re}\left(W\left(z_{e0}\right)\right)=0$. This mode is
developed when the wavelength $\lambda$ meets the inequality
\begin{eqnarray}
\lambda<\lambda_{1lh}=\frac{4k_{y}\rho_{Ti}S_{i}}{A_{1}\left(b_{i}\right)}
\left(1+\tau-G_{1}\right),\label{15}
\end{eqnarray}
where index $1lh$ means the  long-wavelength threshold of the
ion-hydrodynamic mode of the first cyclotron harmonic. The
dispersion of the ion cyclotron waves and the growth rate obtained
from Eq. (\ref{14}) are:
\begin{eqnarray}
\delta\omega=\frac{\delta\omega_{01}}{2},\qquad\gamma=\frac{\delta\omega_{01}}{2}
\sqrt{\frac{\lambda_{1lh}-\lambda}{\lambda}}.\label{16}
\end{eqnarray}
Note, that $\delta\omega$ does not depend on the variable $\lambda$.
The growth rate of the ion-hydrodynamic mode away from the threshold
of stability, as well as for the ion-kinetic mode, varies slightly
with changes of $k_{y}\rho_{Ti}$ for $k_{y}\rho_{Ti}\gtrsim1$, while
for $k_{y}\rho_{Ti}<1$ the growth rate decreases as
$\left(k_{y}\rho_{Ti}\right)^{3/2}$. The development of hydrodynamic
mode of the shear-flow-driven EIC instability will manifest itself
in the turbulent heating of ions across the magnetic field due to
the effect of ion cyclotron resonances broadening.

With a decrease of $\lambda$ the growth rate increases until the
value $\left|z_{i1}\right|$ becomes of order unity, where asymptotic
form of $W$ - function at $\left|z_{i1}\right|>1$ is non-applicable.
As it was shown by numerical estimates, the maximum growth rate
occurs at $\left|z_{i1}\right|\simeq1$, i. e. on the boundary of the
ion-kinetic and ion-hydrodynamic modes of this instability.

When the inequality $\lambda>\lambda_{1lh}$ holds, the ion cyclotron
oscillations split in two hydrodynamically stable modes for which
$\delta\omega$ is equal to
\begin{eqnarray}
\delta\omega=\frac{\delta\omega_{01}}{2}\left(1\pm\sqrt{\frac{\lambda-
\lambda_{1lh}}{\lambda}}\right).\label{17}
\end{eqnarray}
The dispersion with sign plus corresponds to the EIC mode modified
by the flow velocity shear and exists in a plasma without shear
flow. Second mode with sign minus exists only in a plasma with a
shear flow at the same condition as ion-kinetic and ion-hydrodynamic
modes, namely at $\sqrt{2}k_{y}\rho_{Ti}S_{i}>1$. It is the
shear-flow-driven electron-kinetic EIC mode. For the case of a
plasma with no electron current only shear-flow-driven EIC mode may
be unstable due to combined effect of an  inverse electron Landau
damping and velocity shear. The growth rate of this electron-kinetic
EIC mode is approximately \cite{Mikhailenko06}
\begin{eqnarray}
\gamma\approx\omega_{ci}\frac{\sqrt{\pi}\tau
k_{y}^{2}\rho_{Ti}^{2}S_{i}^{2}}
{A_{1}\left(b_{i}\right)\sqrt{2\mu\lambda\left(\lambda-\lambda_{1lh}\right)}}e^{-\lambda^{2}/2\mu}.\label{18}
\end{eqnarray}
The long-wavelength limit of the electron-kinetic mode may be
evaluated from the condition $z_{e0}=\lambda/\sqrt{2\mu}<1$, which
gives the threshold value $\lambda_{1lh}<\sqrt{2\mu}$ which define
the additional condition for the excitation of the electron-kinetic
mode. The estimates performed by Amatucci\cite{Amatucci} for ion
flows propagating in the Earth's ionosphere give the normalized flow
velocity shear $S_{i}\sim0.4$ for hydrogen ions and $S_{i}\sim 6$
for oxygen ions. The conditions of the shear-flow-driven EIC
instability excitation are satisfied for hydrogen ions at
$k_{y}\rho_{Ti}\gtrsim3$ and for oxygen ions at
$k_{y}\rho_{Ti}\gtrsim1$. Then the inequality
$\lambda_{1lh}<\sqrt{2\mu}$ for oxygen ions is met, whereas for
hydrogen ions it is not satisfied. Thus for the conditions in the
Earth's ionosphere the electron-kinetic mode of the
shear-flow-driven EIC instability may be excited only for the flows
of heavy (oxygen) ions. In addition, the development of electron
kinetic mode leads to heating of the electrons along the magnetic
field as well as to the turbulent heating of ions across the
magnetic field.

We numerically solved the dispersion equation (\ref{1}) and obtained
the plots for the dispersion and growth rate depending on the
normalized wavelength for oxygen ions (Fig. 1). The growth rate was
calculated for two cases: with neglecting electron Landau damping
(dashed line) and with accounting of it (dashed-dotted line). The
first curve presents the ion-kinetic and ion-hydrodynamic modes of
the shear-flow-driven EIC instability. One can see that the limiting
wavelengths are equal approximately to 4 and 76 units; these values
are close to theoretical magnitudes, which were calculated from Eqs.
(\ref{5}) and (\ref{15}). The growth rate has maximal value for
$\lambda$ values for which $|z_{1i}|\simeq1$. Electron Landau
damping leads to a decrease of the growth rate of the
ion-hydrodynamic mode, however the waves with the wavelengths
$\lambda>\lambda_{1lh}$ become unstable (second curve). In this
case, we have the electron-kinetic mode of the shear-flow-driven EIC
instability. As it is seen from Fig. 1, the EIC oscillations split
into two modes at $\lambda>\lambda_{1lh}$, that is predicted by Eq.
(\ref{15}).

\begin{figure}
\includegraphics[width=9cm]{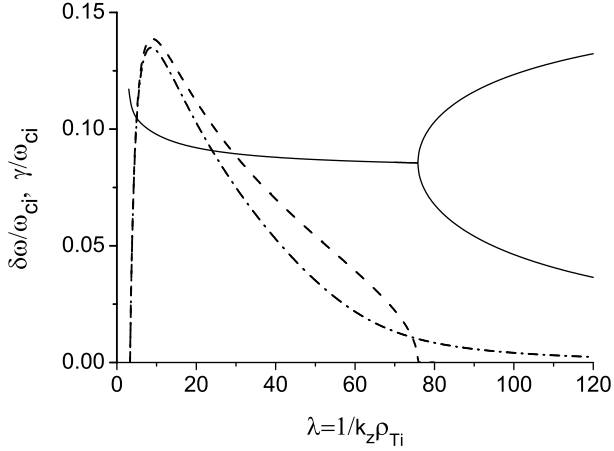}\\
\caption{The dispersion (solid line) and the growth rate (dashed and dashed-dotted lines)
of the EIC oscillations versus the normalized wavelength. Dashed line - the growth rate
with neglecting the Landau damping by electrons, dashed-dotted line - the growth rate
accounted for the Landau damping by electrons. $S_{i}=3, k_{\perp}\rho_{Ti}=1$.}\label{1}
\end{figure}

So, the first harmonic of the shear flow-driven EIC instability can
be be considered as a set of three modes,  the ion-kinetic mode, the
ion-hydrodynamic mode and the electron-kinetic mode. Each of them
exists in a certain ranges of wavelengths along the magnetic field,
however they have common signatures such as condition
$\sqrt{2}k_{y}\rho_{Ti}S_{i}>1$ for their excitation and the growth
of the growth rate with an increase of $k_{y}\rho_{Ti}$.

\section{THE INSTABILITY OF THE HIGH CYCLOTRON HARMONIC}

Now we investigate the instability of high cyclotron harmonic,
$\omega
\left(\mathbf{k}\right)=n'\omega_{ci}+k_{z}V_{0i}+\delta\omega
\left(\mathbf{k}\right)$ with $|n'|\geq 2$ and
$\delta\omega\left(\mathbf{k}\right)\ll \omega_{ci}$. Assume that
$z_{in}$ in the sum over cyclotron harmonics has an arbitrary value
for the $n=n'$ term, while in remaining sum $|z_{in}|>1$, for which
the asymptotic form of $W$ - function for large argument values may
be used. The summation over cyclotron harmonics at
$k_{y}\rho_{Ti}\gg 1$ gives approximately
\begin{eqnarray}
\sum\limits_{n \neq n'}W\left(z_{i
n}\right)A_{n}\left(b_{i}\right)\left(\frac{\omega-k_{z}V_{0i}}
{\sqrt{2}k_{z}v_{Ti}}-\frac{k_{y}}{k_{z}}S_{i}z_{in}\right)\approx
\psi\left(z_{\perp}\right)
-\frac{k_{y}}{k_{z}}S_{i}\left(1-A_{n'}\left(b_{i}\right)\right),\label{19}
\end{eqnarray}
where
$z_{\perp}=\left(n'\omega_{ci}+\delta\omega\right)/\sqrt{2}k_{y}
v_{Ti}\approx n'/k_{y}\rho_{Ti}$ and $\psi\left(z_{\perp}\right)
=-2z_{\perp}e^{-z_{\perp}^{2}}\int\limits_{0}^{z_{\perp}}e^{t^{2}}dt$.
Then dispersion equation (\ref{3}) can be presented in the form
\begin{eqnarray}
k^2\lambda_{Di}^2\varepsilon\left({\mathbf{k},\omega
}\right)&&=1+\tau\left(1+i\sqrt{\pi}z_{e0}W\left(z_{e0}\right)\right)+\psi\left(z_{\perp}\right)
-\frac{k_{y}}{k_{z}}S_{i}A_{n'}\left(b_{i}\right)\nonumber\\
&& +i\sqrt{\pi}W\left(z_{i
n'}\right)A_{n'}\left(b_{i}\right)\left(\frac{n'\omega_{ci}}
{\sqrt{2}k_{z}v_{Ti}}+z_{in'}-\frac{k_{y}}{k_{z}}S_{i}z_{in'}\right)=0.\label{20}
\end{eqnarray}

Consider first the ion-kinetic mode of the shear-flow-driven EIC
instability.  The approximate instability threshold at the short
wavelength limit along the magnetic field can be found from Eq.
(\ref{20}). Assuming that the electrons are adiabatic we equate to
zero the real and imaginary parts of the Eq. (\ref{20}), and obtain
the set of equations for variables $\lambda$ and $z_{in'}$, which is
similar to the corresponding set for the first cyclotron harmonic
(\ref{4})
\begin{eqnarray}
\left\{\begin{array}{ll}n'\lambda/\sqrt{2}+z_{in'}-\lambda k_{y}\rho_{Ti}
S_{i}z_{in'}=0,
\\ 1+\tau+\psi\left(z_{\perp}\right)
-\lambda k_{y}\rho_{Ti}S_{i}A_{n'}\left(b_{i}\right)=0.\end{array} \right.\label{21}
\end{eqnarray}
This set has a solution when the inequality $k_{y}\rho_{Ti}S_{i}>0$
is met. For this case we obtain the short-wavelength threshold value
$\lambda_{n's}$ for the excitation of the instability, at which
normalized complex frequency $z_{n's}$ becomes real,
\begin{eqnarray}
\lambda_{n's}=\frac{1}{k_{y}\rho_{Ti}S_{i}A_{n'}\left(b_{i}\right)}\left(1+
\tau+\psi\left(z_{\perp}\right)\right),\label{22}
\end{eqnarray}
\begin{eqnarray}
z_{n's}=\frac{n'}{\sqrt{2}S_{i}k_{y}\rho_{Ti}}\left(1+\frac{A_{n'}\left(b_{i}
\right)}{1-G_{n'}+\tau}\right),\label{23}
\end{eqnarray}
where
$G_{n'}=
A_{n'}\left(b_{i}\right)-\psi\left(z_{\perp}\right)$ and index $n's$ means the  short-wavelength  threshold of
instability for $n'$-th cyclotron harmonic.  The value
$\delta\omega=\delta\omega_{0n'}$ at the instability threshold is
\begin{eqnarray}
\delta\omega_{0n'}=\frac{n'\omega_{ci}A_{n'}\left(b_{i}\right)}{1-G_{n'}+\tau}.\label{24}
\end{eqnarray}
The solution of Eq. (\ref{20}) at the vicinity of stability
threshold can be obtained by use the same approach as for the first
cyclotron harmonic. This yields the dispersive part and the growth
rate for ion cyclotron instability,
\begin{eqnarray}
\delta\omega\simeq\delta\omega_{0n'}\frac{\lambda_{n's}}{\lambda}+\delta\omega_{0n'}
\left(\frac{\sqrt{2}k_{y}\rho_{Ti}S_{i}\text{Im}W\left(z_{
n's}\right)}{\sqrt{\pi}n'|W\left(z_{
n's}\right)|^{2}}-\frac{A_{n'}\left(b_{i}\right)}{1-G_{n'}+\tau}\right)\left
(1-\frac{\lambda_{n's}}{\lambda}\right),\label{25}
\end{eqnarray}
\begin{eqnarray}
\gamma\simeq\delta\omega_{0n'}\frac{\sqrt{2}k_{y}\rho_{Ti}S_{i}\text{Re}W\left(z_{
n's}\right)}{\sqrt{\pi}n'|W\left(z_{
n's}\right)|^{2}}\left(1-\frac{\lambda_{n's}}{\lambda}\right).\label{26}
\end{eqnarray}
The growth rate is affected by threshold value $z_{n's}\approx
n'/\sqrt{2}k_{y}\rho_{Ti}S_{i}$, because
$\gamma\propto\text{Re}W\left(z_{n's}\right)=exp\left(-z_{n's}^{2}\right)$.
The growth rate of instability is exponentially small when the flow
velocity shear and the transverse wave number are such that
$\sqrt{2}k_{y}\rho_{Ti}S_{i}<n'$, whereas, at opposite inequality
the growth rate  is not exponentially small. The magnitude of the
growth rate is also affected by the factor
$k_{y}\rho_{Ti}A_{n'}\left(b_{i}\right)$. The function
$A_{n'}\left(b_{i}\right)$ at $k_{y}\rho_{Ti}\thicksim n'\gg1$ has
asymptotic form
 \begin{eqnarray}
A_{n'}\left(b_{i}\right)
 \thicksim\left(1/\sqrt{2\pi} k_{y}\rho_{Ti}\right)\exp\left(-n'^{2}/2k_{y}^{2}
 \rho_{Ti}^{2}\right),\label{27}
\end{eqnarray}
which implies that long waves with $k_{y}\rho_{Ti}<n'$ have
exponentially small growth rate. Thus, waves with longitudinal
wavelength $\lambda>\lambda_{n's}$ and transverse wavenumbers such
as $k_{y}\rho_{Ti}\gtrsim n'$ and $\sqrt{2}k_{y}\rho_{Ti}S_{i}>n'$
are unstable. We note also that the threshold wavelength
$\lambda_{n's}$ (\ref{22}) with these transverse wavenumbers is
approximately equal to the corresponding magnitude for the first
cyclotron harmonic (\ref{5}) with transverse wavenumbers
$k_{y}\rho_{Ti}\gtrsim 1$.

As it is for the fundamental ion cyclotron harmonic, the growth rate
of the ion-kinetic mode of the shear-flow-driven EIC instability
increases with an increase of wavelength along the magnetic field.
Simultaneously the value $\left|z_{in'}\right|$ increases also. When
the transverse wave numbers meet the inequality
$\sqrt{2}k_{y}\rho_{Ti}S_{i}<n'$, the growth rate, remaining
exponentially small quantity, reaches a maximum at a certain value
of $\lambda$ and then rapidly decreases. When
$\sqrt{2}k_{y}\rho_{Ti}S_{i}>n'$, the ion-kinetic mode with the
fulfilment of condition $\left|z_{in'}\right|>1$ turn into the
ion-hydrodynamic mode. The dispersion equation for the
ion-hydrodynamic mode of the shear-flow-driven EIC instability can
be obtained from Eq. (\ref{20}) by using the asymptotic form of $W$
- function for large argument values
\begin{eqnarray}
1+\tau\left(1+i\sqrt{\pi}z_{e0}W\left(z_{e0}\right)\right)-G_{n'}-
\frac{A_{n'}\left(b_{i}\right)}{\sqrt{2}z_{in'}}\lambda+
\frac{A_{n'}\left(b_{i}\right)}{2z_{in'}^{2}}k_{y}\rho_{Ti}S_{i}\lambda=0.\label{28}
\end{eqnarray}
where $z_{e0}=n'\lambda/\sqrt{2\mu}$. The solution of Eq. (\ref{28}) is
\begin{eqnarray}
&&z_{in'}=\left[\lambda
A_{n'}\left(b_{i}\right)\pm\sqrt{\left(\lambda
A_{n'}\left(b_{i}\right)\right)^{2}-4\left(1+\tau\left(1+i\sqrt{\pi}z_{e0}
W\left(z_{e0}\right)\right)-G_{n'}\right)S_{i}\lambda
k_{y}\rho_{Ti}A_{n'}\left(b_{i}\right)}\right]
\nonumber\\
&&\times\left[2\sqrt{2}k_{y}\rho_{Ti}\left
(1+\tau\left(1+i\sqrt{\pi}z_{e0}W\left(z_{e0}\right)\right)-G_{n'}\right)\right]^{-1}.\label{29}
\end{eqnarray}
The value $z_{e0}$ may be more or less than the unity depending on
the number of cyclotron harmonic. We find from Eq. (\ref{29}) the
long wavelength threshold for $n'$-th harmonic of the
ion-hydrodynamic mode, assuming that $z_{e0}\gg1$. The instability
develops when the wavelength $\lambda$ meets the inequality
\begin{eqnarray}
\lambda<\lambda_{n'lh}=\frac{4k_{y}\rho_{Ti}S_{i}}{n'^{2}A_{n'}\left(b_{i}\right)}
\left(1-G_{n'}\right),\label{30}
\end{eqnarray}
where index $n'lh$ means the long-wavelength threshold for the
$n'$-th cyclotron harmonic of the ion-hydrodynamic mode. For the
transverse wavenumbers $k_{y}\rho_{Ti}\gtrsim n'$, the
long-wavelength threshold (\ref{30}) is equal approximately to the
corresponding value for the first cyclotron harmonic (\ref{16}) with
transverse wavenumbers $k_{y}\rho_{Ti}\gtrsim 1$. The dispersion and
the growth rate of EIC oscillations obtained from (\ref{29}) are:
\begin{eqnarray}
\delta\omega=\frac{\delta\omega_{0n'}}{2},\qquad
\gamma=\frac{\delta\omega_{0n'}}{2}\sqrt{\frac{\lambda_{n'lh}-\lambda}{\lambda}}\label{31}
\end{eqnarray}
which is equal approximately to the corresponding values, obtained
above for the first cyclotron harmonic. The growth rate (\ref{31})
far from the stability threshold, as well as growth rate (\ref{26})
for the kinetic mode, varies slightly with changing $k_{y}\rho_{Ti}$
for $k_{y}\rho_{Ti}>n'$, while for opposite inequality it decreases
exponentially with decreasing of $k_{y}\rho_{Ti}$. With decreasing
of $\lambda$, the growth rate increases until the value
$\left|z_{in'}\right|$ becomes of the order of unity, for which the
asymptotic form of $W$ - function, obtained for
$\left|z_{in'}\right|>1$ becomes unapplicable. As it was shown by
numerical estimates, the growth rate has a  maximum at
$\left|z_{in'}\right|\thicksim1$, i. e. on the boundary of the
ion-kinetic and ion-hydrodynamic modes.

When the inequality $\lambda>\lambda_{n'lh}$ holds, the growth rate
of EIC ion-hydrodynamic mode is zero. In this case, the ion
cyclotron oscillations  split onto two hydrodynamically stable EIC
modes, namely, current-driven modified by shear and
shear-flow-driven modes. For these modes
\begin{eqnarray}
\delta\omega=\frac{\delta\omega_{0n'}}{2}\left(1\pm\sqrt{\frac{\lambda-
\lambda_{n'l}}{\lambda}}\right).\label{32}
\end{eqnarray}
For the excitation of the electron-kinetic shear-flow-driven EIC
instability in plasma without electron current, the inequality
$z_{e0}=n'\lambda/\sqrt{2\mu}<1$ with $\lambda>\lambda_{n'lh}$ must
hold. Hence, the restriction on the numbers of unstable cyclotron
harmonics can be obtained, $n'<\sqrt{2\mu}/\lambda_{n'lh}$. The last
inequality may be satisfied only for heavy ions; specifically, for
the oxygen ions we have $n'<6$.

Note, that the ion-hydrodynamic mode does not excite for values of
the flow velocity shear and transverse wave numbers, for which
$\sqrt{2}k_{y}\rho_{Ti}S_{i}<n'$; the splitting of EIC oscillations
on two modes also is absent in that case. Thus, a necessary
condition for the excitation of high $n'$-th harmonics of
shear-flow-driven EIC instability is determined by the inequalities
$\sqrt{2}k_{y}\rho_{Ti}S_{i}>n'$ and $k_{y}\rho_{Ti}\gtrsim n'$
which are similar to the corresponding conditions for the
fundamental cyclotron harmonic.

\section{CONCLUSIONS}
In this paper, we have studied the EIC instabilities driven by the
velocity shear of plasma flow along the magnetic field (the
shear-flow-driven EIC instability) for different values of the
wavelength along the magnetic field. The EIC oscillations with first
$n=1$ and high $|n|\geq 2$ cyclotron harmonics are considered. It
was shown that the shear-flow-driven EIC instabilities, depending on
the wavelengths along the magnetic field, are of three types: the
ion-kinetic mode, the ion-hydrodynamic mode and the electron-kinetic
mode. Each of these modes are caused by different mechanisms.

The ion-kinetic mode is the most short-wavelength among them. It
develops as a result of the combined effect of the inverse of ion
cyclotron damping and leads to the ions heating along the magnetic
field. The short-wavelength limit along the magnetic field
$\lambda_{ns}$ for the ion-kinetic  shear-flow-driven EIC
instability for $n=1$ and $|n|\geq 2$ cyclotron harmonics were
determined by Eqs. (\ref{5}) and (\ref{22}),respectively, which are
almost equal. The ion-hydrodynamic mode is dominant for longer
wavelength of ion cyclotron waves along the magnetic field. The
longitudinal wavelength $\lambda_{bn}$, which conventionally
separates these modes is determined by equality
$|z_{in}|=\lambda_{bn}|\omega-n\omega_{ci}-k_{z}V_{0i}|/\sqrt{2}\omega_{ci}
\simeq1$ and the magnitudes $\lambda_{bn}$ for $n=1$ and $|n|\geq 2$
cyclotron harmonics are approximately equal. Thus, the ion-kinetic
mode is excited at wavelengths
$\lambda_{ns}<\lambda\lesssim\lambda_{bn}$. We have numerically
obtain that the growth rate of the shear-flow-driven EIC instability
has a maximal value at the boundary of ion-kinetic and
ion-hydrodynamic modes, i. e. at $\lambda\simeq\lambda_{bn}$.

The ion-hydrodynamic shear-flow-driven EIC instability has the
similar mechanism of development, as for the aperiodic instability
driven by  velocity  shear, which was discovered by
D'Angelo\cite{Angelo}. It exists in the range of wavelengths along
the magnetic field, determined by inequality $|z_{in}|>1$. The
dispersion as well as the growth rate of the shear-flow-driven EIC
instability are approximately the same for all cyclotron harmonics.
The ion-hydrodynamic mode has a limited value of the long wavelength
along the magnetic field $\lambda_{nlh}$, which is determined by
Eqs. (\ref{15}) and (\ref{30}) for $n=1$ and $|n|\geq 2$ cyclotron
harmonics respectively; in so doing for the transverse wave numbers
$k_{y}\rho_{Ti}\gtrsim n$ the boundary wavelengths $\lambda_{nlh}$
for different cyclotron harmonics are approximately equal. Thus the
ion-hydrodynamic mode is excited in wavelengths range
$\lambda_{bn}\lesssim\lambda<\lambda_{nlh}$. This mode leads to the
turbulent heating of ions across the magnetic field due to the
effect of ion cyclotron resonances broadening.

For the wavelengthes $\lambda>\lambda_{nlh}$, the EIC oscillations
split into two hydrodynamically stable modes with the dispersion
evaluated by Eqs. (\ref{17}) and (\ref{32}) for $n=1$ and $|n|\geq
2$ cyclotron harmonics respectively. One of them with sign minus in
Eqs. (\ref{17}) and (\ref{32}) is the electron-kinetic mode of the
shear-flow-driven EIC instability It becomes unstable even in
currentless plasma flow due to an inverse of electron Landau damping
when $z_{e0}=n\lambda/\sqrt{2\mu}\lesssim1$. The excitation of
electron-kinetic mode depends on the ions mass and of cyclotron
harmonic mode number. We have found that it may be excited only for
heavy ions, such as oxygen ions, while for a hydrogen ions this mode
is not excited. Then the range of wavelength of the electron-kinetic
mode is evaluated by inequality
$\lambda_{nhl}<\lambda\lesssim\lambda_{nl}\simeq\sqrt{2\mu}/n$.

We have obtained the conditions for the excitation of the
shear-flow-driven EIC instabilities, which connects the normalized
shear and wave numbers transversely the magnetic field. These
conditions are identical for all considered modes, and determined as
$\sqrt{2}k_{y}\rho_{Ti}S_{i}>n$. When this condition holds, all
modes excited simultaneously.These instabilities result in the ion
heating both along and across the magnetic field as well as in the
electron heating along the magnetic field.

\end{document}